\documentclass[5p,twocolumn,number]{elsarticle}

\usepackage{lineno,hyperref}
\modulolinenumbers[5]

\usepackage{graphicx}
\usepackage{xcolor}
\usepackage{amsmath}
\usepackage{textcomp}
\journal{NIMA}

\begin{document}

\begin{frontmatter}

\title{A Liquid Hydrogen Target for the MUSE Experiment at PSI}

\author[UMICH]{P.~Roy}
\author[UMICH]{S.~Corsetti }
\author[UMICH]{M.~Dimond }
\author[UMICH]{M.~Kim }
\author[UMICH]{L.~Le Pottier }
\author[UMICH]{W.~Lorenzon\corref{mycorrespondingauthor}}
\author[UMICH]{R.~Raymond }
\author[UMICH]{H.~Reid }
\author[UMICH]{N.~Steinberg }
\author[UMICH]{N.~Wuerfel }
\author[PSI]{K.~Deiters }
\author[GW]{W.J.~Briscoe }
\author[GW]{A.~Golossanov }
\author[RUT,PSI]{T. Rostomyan }

\cortext[mycorrespondingauthor]{Corresponding author}

\address[UMICH]{Randall Laboratory of Physics, University of Michigan, Ann Arbor, MI 48109-1040, USA}
\address[PSI]{Paul Scherrer Institut, CH-5232 Villigen, PSI, Switzerland}
\address[GW]{Department of Physics, The George Washington University, Washington, D.C. 20052, USA}
\address[RUT]{Rutgers, The State University of New Jersey, Piscataway, New Jersey 08855, USA}

\begin{abstract}
A 280\,ml liquid hydrogen target has been constructed and tested for the MUSE experiment at PSI to investigate the proton charge radius via simultaneous measurement of elastic muon-proton and elastic electron-proton scattering. To control systematic uncertainties at a sub-percent level, strong constraints were put on the amount of material surrounding the target and on its temperature stability. The target cell wall is made of $120\,\mu$m-thick Kapton$^{\text \textregistered}$, while the beam entrance and exit windows are made of $125\,\mu$m-thick aluminized Kapton$^{\text \textregistered}$. The side exit windows are made of Mylar$^{\text \textregistered}$ laminated on aramid fabric with an areal density of $368$\,g/m$^2$. The target system was successfully operated during a commissioning run at PSI at the end of 2018. The target temperature was stable at the 0.01\,K level. This suggests a density stability at the 0.02\% level, which is about a  factor of ten better than required.
\end{abstract}

\begin{keyword}
\texttt{Liquid hydrogen target\sep muon beam\sep MUSE \sep elastic scattering}\\

\end{keyword}

\end{frontmatter}

%\linenumbers

\section{Introduction}

Until recently, the proton charge radius had been well established at $r_p=0.8768(69)$\,fm~\cite{CODATA2006} from electronic-hydrogen spectroscopy and electron-proton scattering measurements. However in 2010, a novel experiment using muonic-hydrogen spectroscopy~\cite{Pohl2010} determined the proton charge radius to be $r_p= 0.84184(67)$\,fm. This roughly five-standard-deviation difference has become known as the proton radius puzzle~\cite{protonpuzzle-2013}.

The MUon Scattering Experiment (MUSE)~\cite{MUSE_TDR}, located in the PiM1 area of the Paul Scherrer Institute (PSI) in Switzerland, is part of a suite of experiments that aim to resolve the proton radius puzzle. MUSE attempts to determine the proton charge radius through simultaneous measurements of muon-proton and electron-proton elastic scattering cross sections with high precision. The experimental kinematics cover three beam momenta of about 115, 153, and 210\,MeV/$c$ and scattering angles in the range of $20^{\circ} - 100^{\circ}$, corresponding to $Q^2$ of approximately 0.002\,(GeV/c)$^2 - 0.08\,$(GeV/c)$^2$, the range of the form factor with greatest sensitivity to the radius~\cite{MUSE_TDR}.

At the heart of the experiment is a liquid hydrogen (LH$_2$) target that needs to provide stable density and sufficient cooling power to minimize uncertainty in target length in order to allow cross section measurements at the sub-percent level. Since hydrogen is highly explosive, special precautions had to be taken to ensure safe handling. Driven by the science needs for MUSE, the experimental requirements for the target system~\cite{MUSE_TDR} include not only a LH$_2$ cell, but also an empty cell for studying background generated from the target walls and a thin, solid target for precision vertex reconstruction and detector alignment. To facilitate rapid switching between the LH$_2$ and empty target cells, and to accommodate the MUSE spectrometer geometry, the target system has to be movable in the vertical direction. Furthermore, large thin vacuum windows on either side of the beam are needed to cover the very large solid angle subtended by the detectors while introducing minimal amounts of multiple scattering.

A LH$_2$ target system that met all experimental and safety requirements was designed and fabricated in close collaboration with Creare Inc, and then installed and thoroughly tested at PSI. This paper describes the major components of the target apparatus and presents data demonstrating the successful operation of the LH$_2$ target.

\begin{figure*}[t!]
	\centering
	\includegraphics[width=0.69\textwidth]{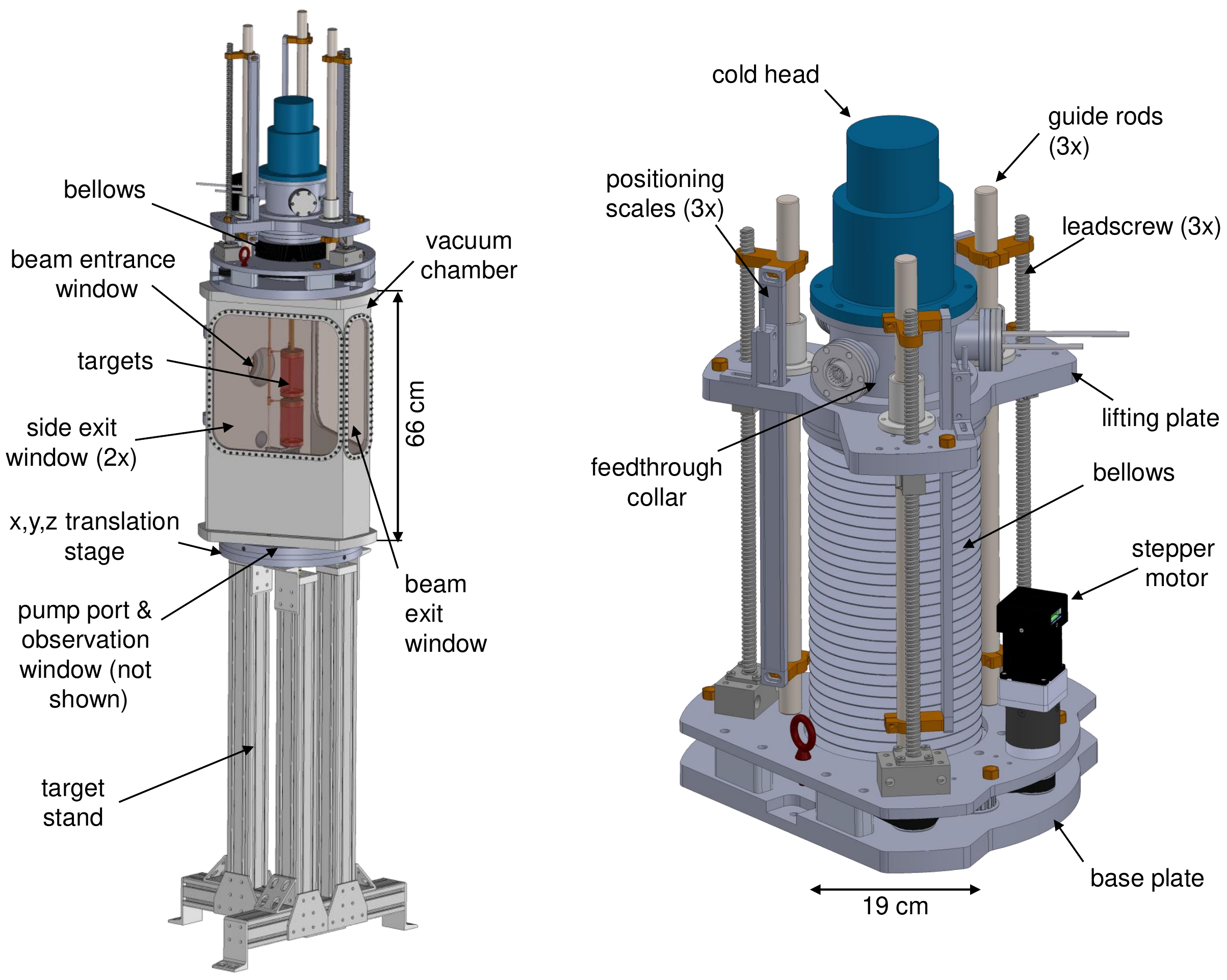}
	\caption{Left Panel: Schematic view of the trapezoidal vacuum chamber and the lifting assembly mounted on a support stand. The vacuum chamber is 66\,cm high. Right Panel: Close-up view of the lifting assembly components  that facilitate the vertical motion of the target ladder. The bellows outer diameter is 19\,cm.}
	\label{Vac_cham}
\end{figure*}

\section{Target System}
\label{sec:Target_app}

The MUSE target system, shown in Fig.~\ref{Vac_cham}, consists of a target ladder with a variety of targets, a cryocooler and condenser assembly that liquefies the hydrogen gas, a vacuum chamber that houses the target ladder, a lifting mechanism that allows fast switching of the target cells, a gas handling system, and a slow control system that regulates temperature and the hydrogen gas flow in and out of the target cell.

\subsection{Target ladder}
\begin{figure}[ht!]
	\centering
	\includegraphics[width=0.4\textwidth]{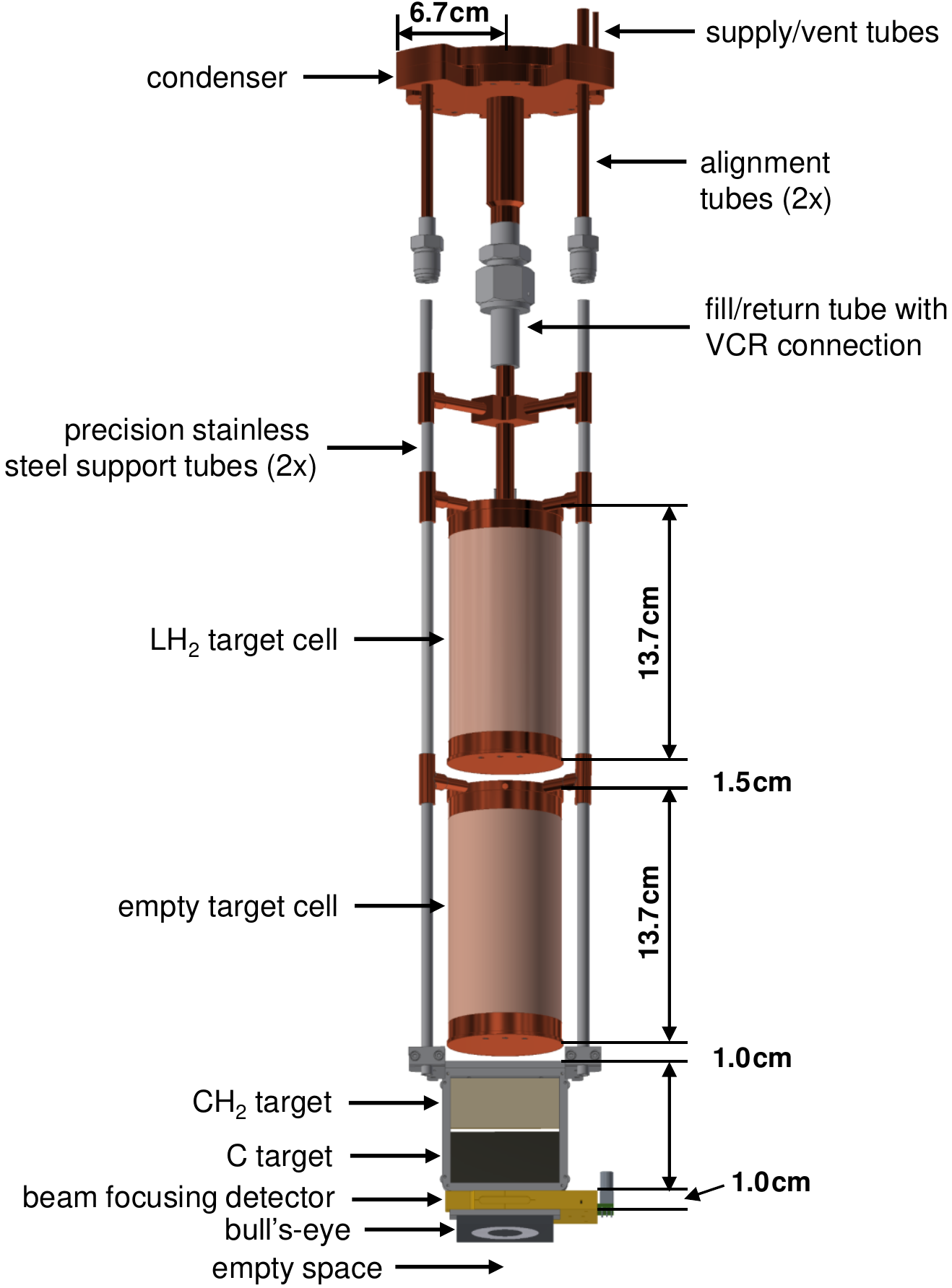}
	\caption{CAD drawing of the target ladder connected to the condenser showing the target positions (LH$_2$ cell, empty cell, CH$_2$ target, C target, empty space) as well as the location of the beam focusing detector and the optical ``bull's-eye'' target. A total vertical displacement of 34\,cm is required to attain all positions.  Note that the superinsulation wrap and the copper braids are omitted for clarity.}
	\label{Tar_Lad}
\end{figure}

\begin{figure}[ht!]
	\centering
	\includegraphics[width=0.5\columnwidth]{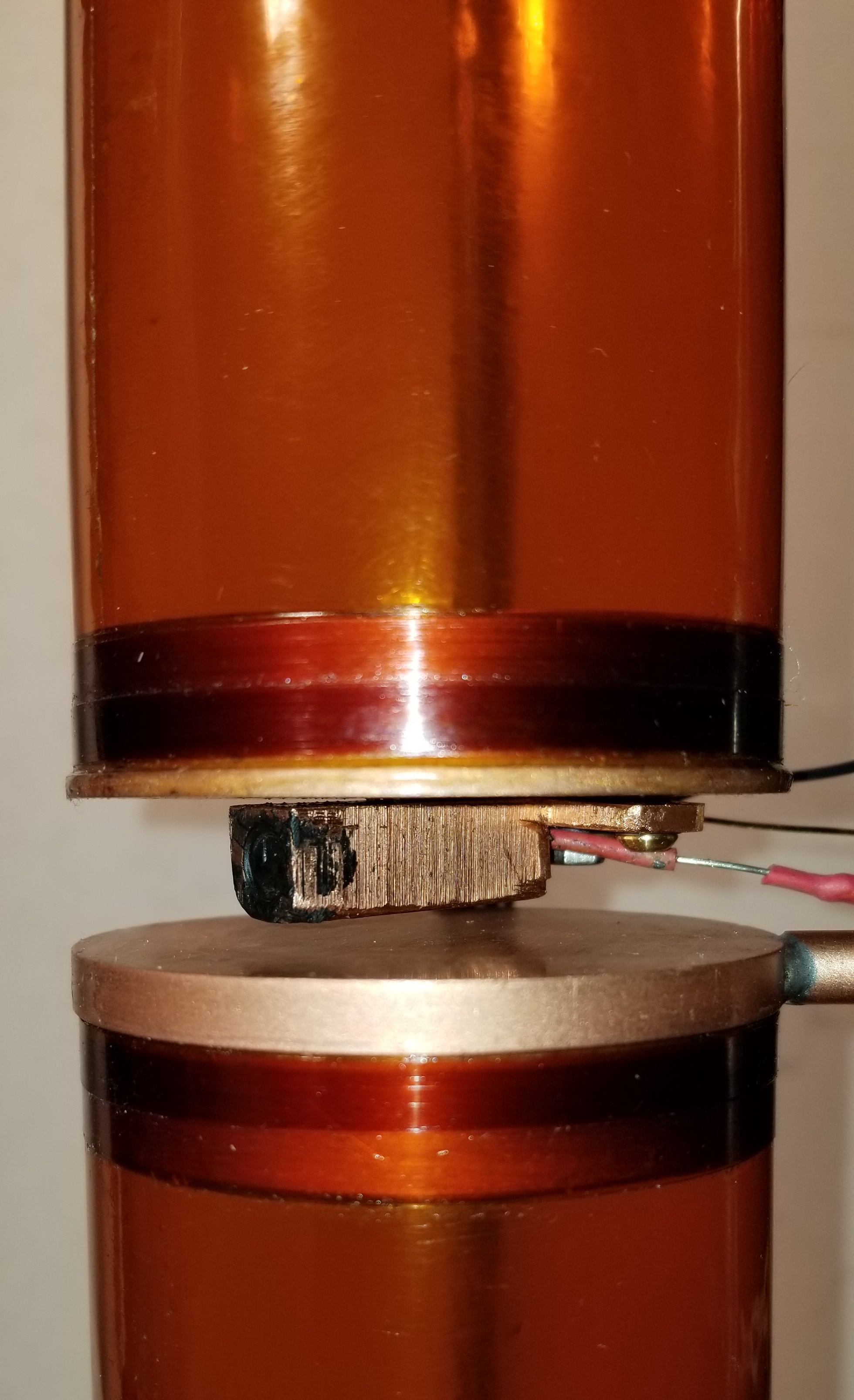}
	\caption{A photograph of the LH$_2$ cell (top) and empty cell (bottom). The cell is reinforced with two wide and two narrow Kapton$^{\text \textregistered}$ strips glued to the end caps. Also shown is the cartridge heater, containing one 50\,$\Omega$, 50\,W resistor, and one Cernox$^{\text \textregistered}$ temperature sensor (partially hidden) attached to the end cap of the LH$_2$ cell.}
	\label{LH2_cell}
\end{figure}

The target ladder, shown in Fig.~\ref{Tar_Lad}, can be moved in and out of the particle beam in the vertical direction. The main target, shown in Fig.~\ref{LH2_cell}, is a $280$\,ml LH$_2$ cell assembled from a $6$\,cm inner diameter Kapton$^{\text \textregistered}$ cylinder and copper end caps that provide strength and stability to the Kapton$^{\text \textregistered}$ wall.
It is operated at a temperature of 20.65\,K and a pressure of 1.1\,bar.

The cell wall is made with four wraps of a single sheet of $25\,\mu$m-thick Kapton$^{\text \textregistered}$ foil which is 130\,mm wide and 758\,mm long. The four wraps are glued together using Stycast 1266 epoxy. The thickness of the epoxy layers is controlled to achieve a total cell wall thickness of about $120\,\mu$m. To form a strong glue joint between the end caps and the cylinder, each copper end cap reaches 1\,cm into the Kapton$^{\text \textregistered}$ cylinder. This leads to a cell with a $11$\,cm high, thin Kapton$^{\text \textregistered}$ wall to minimize background from scattering from the copper end caps into the detectors. The cells are further reinforced by first gluing two 9\,mm-wide Kapton$^{\text \textregistered}$ strips, followed by two 5\,mm-wide Kapton$^{\text \textregistered}$ strips over the glue joints between the cylinder and the end caps, as shown in Fig.~\ref{LH2_cell}. Burst tests performed at room temperature have consistently shown that the cells can withstand pressures of over 3.8\,bar. This provides a higher safety factor than the factor of three required by the PSI safety group.

The target cell has two level sensors (with one serving as a backup) mounted on the inner surface of the top end cap. Each sensor is a $100\,\Omega$ Allen Bradley carbon resistor driven at 20\,mA. One Lakeshore Cernox$^{\text \textregistered}$ thin film resistance cryogenic temperature sensor and one ($50\,\Omega$, 50\,W) cartridge heater are mounted on the outer face of the bottom end cap (see Fig.~\ref{LH2_cell}). The temperature sensor, the level sensors, and the heater are all monitored and controlled by the slow control system described in Sec.~\ref{sec:slow_control}. Below the LH$_2$ target cell is an empty target cell which is identical to the LH$_2$ cell. It is used to subtract background originating from the Kapton$^{\text \textregistered}$ walls, and to a lesser extent from the copper end caps of the LH$_2$ cell. A small orifice on the top end cap of the empty cell connects its inner space to the vacuum that surrounds the target ladder. %Thermal braids are soldered to the copper tees of the LH$_2$ and empty cells to establish thermal contact.

\begin{figure}[ht!]
	\centering
	\includegraphics[width=0.4\textwidth]{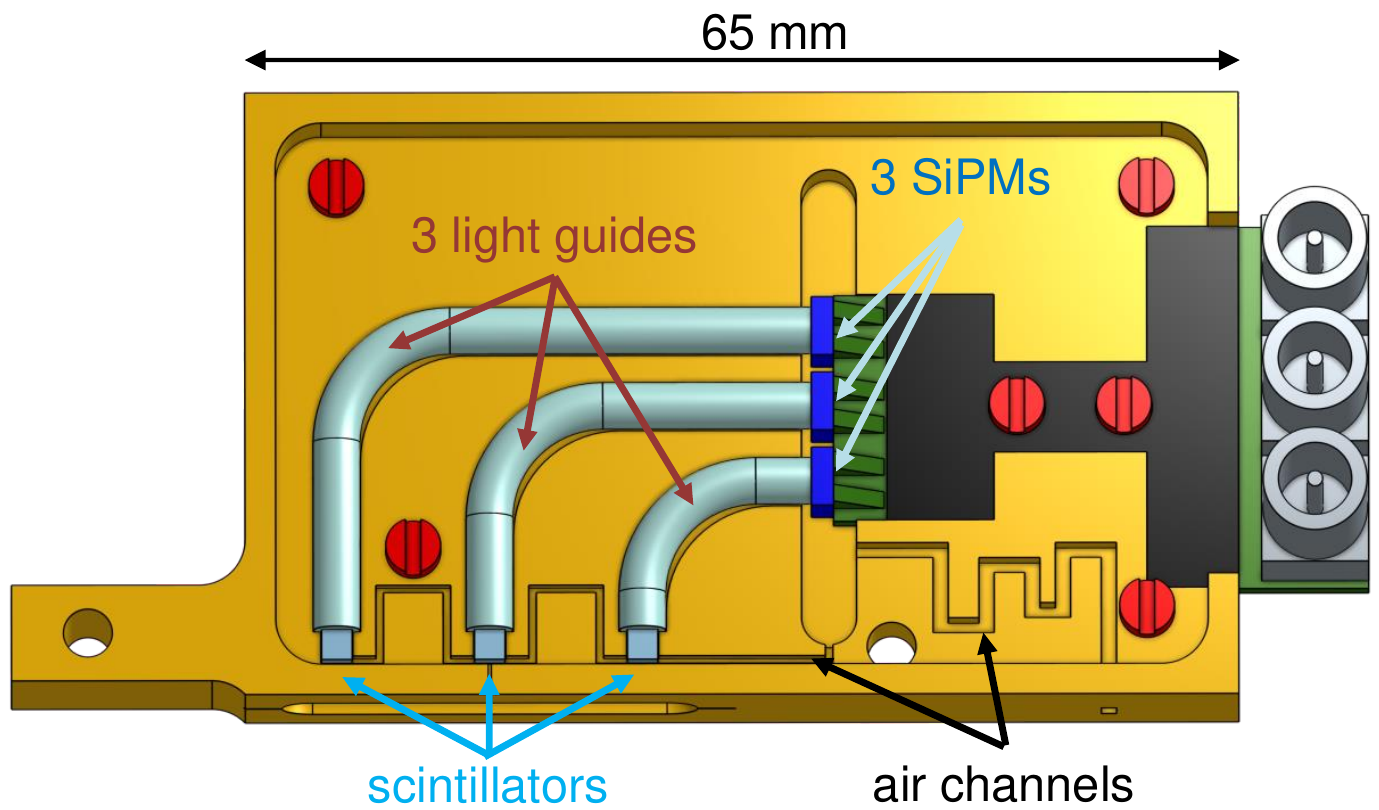}
	\caption{A schematic view of the beam focusing detector. Beam positions are measured with three small scintillators that are connected by light guides to SiPMs.}
	\label{beam_focusing_detector}
\end{figure}

A 1\,mm-thick solid target is attached below the empty cell to allow for precise vertex reconstruction, and to aid in detector alignment. It is split into two parts, with the polyethylene (CH$_2$) target mounted above the carbon (C) target.  A beam focusing detector, installed just below the carbon target, is used to check beam focusing in the target region. The beam focusing detector, shown in Fig.~\ref{beam_focusing_detector}, consists of three scintillators mounted in a horizontal line, with 1\,cm separation. The scintillators are made of $2\times 2\times 2\,$mm$^3$ plastic, and are connected by light guides to silicon photomultipliers (SiPMs)~\cite{BFD}. The beam focusing detector can also be used to map out the vertical beam profile by moving the assembly in the vertical direction.

Just below the beam focusing detector is an optical target. This target is viewed by a camera to determine the horizontal position of the target ladder inside the vacuum chamber to better than 0.1\,mm, using pattern recognition. The lowest target position, noted as ``empty space'' position in Fig.~\ref{Tar_Lad} is used to study background events arising from interactions with materials in the experimental setup aside from the target cell.

The target ladder is connected to the condenser through the hydrogen fill-and-return tube which contains VCR$^{\text \textregistered}$ metal gasket face seal fittings for ease of changing ladders. Two thin tubes, made of precision stainless steel, support  the empty cell, the carbon target and the beam focusing detector. They are positioned so as to minimize interference with scattered particles. The support tubes also maintain alignment between the LH$_2$ cell and the empty cell to within a few hundred microns. Cooling of the empty cell is achieved through copper braids connected to the LH$_2$ cell.

The heat load on the target from the particle beam~\cite{MUSE_TDR} is less than $10\,\mu$W. The approximately 1.4\,W radiation heat load from the vacuum chamber on the target cells is reduced to about $130$\,mW by wrapping $10$ layers of aluminized Mylar$^{\text \textregistered}$ superinsulation around the target ladder and condenser. Heat load on the target due to conduction is negligible.

\subsection{Cryocooler and condenser assembly}
\label{sec:condenser}

The CH110-LT single-stage cryocooler from Sumitomo Heavy Industries Ltd was selected for refrigeration.\footnote{Sumitomo was chosen over Cryomech partly because similar cryocoolers are commonly used at PSI and because Sumitomo has a service center in Darmstadt, Germany.} It has a cooling power of $25$\,W at $20$\,K which is sufficient to cool down and fill the target cell in approximately 2.6\,hours.

\begin{figure}[ht!]
	\centering
	\includegraphics[width=0.95\columnwidth]{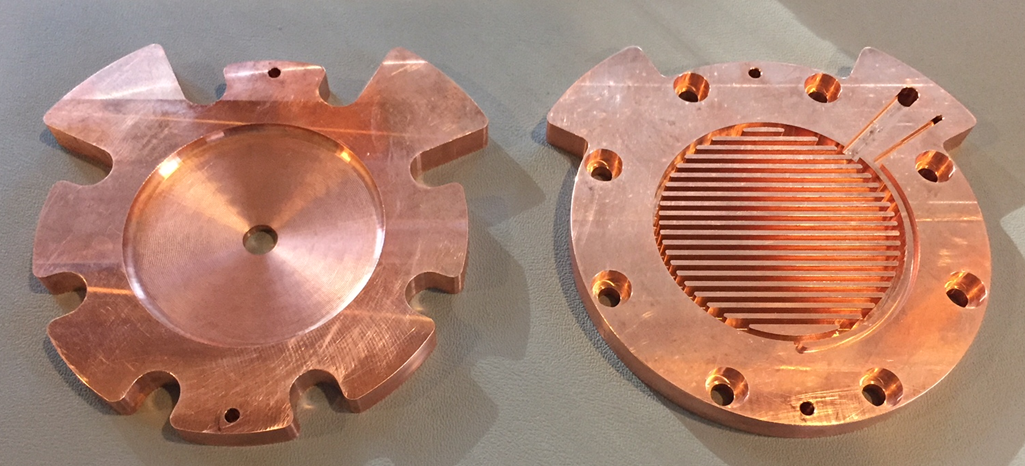}
	\caption{A photograph of the bottom (left) and top (right) condenser plates. The top plate has a series of copper fins to provide a large surface area for condensing hydrogen.}
	\label{condenser}
\end{figure}

The condenser, shown in Fig.~\ref{condenser},
is made of two copper plates soldered together to enclose a cavity with an internal surface area of $346$\,cm$^2$ for collecting and condensing hydrogen gas. The top plate has a series of copper fins to provide the large surface area for condensing hydrogen, while the bottom plate contains a shallow, cone-shaped floor that guides the liquefied hydrogen to the fill tube in the center of the bottom plate. The fill tube is made of reinforced copper with a 8\,mm inner and a 20\,mm outer diameter and forms the sole connection between the condenser and the target ladder. Boiled-off hydrogen from the target cell returns to the condenser cavity through the same tube.

The condenser is bolted to the cryocooler cold head. Apiezon N, a very low vapor pressure vacuum grease, is used to establish good thermal contact between the cold head and the condenser. In order to maintain constant temperature in the target cell, the condenser has two identical $25\,\Omega$, 100\,W heater circuits and two Cernox$^{\text \textregistered}$  temperature sensors which are controlled by the slow control system (see Sec.~\ref{sec:slow_control}). One heater circuit and one temperature sensor are for regular operation while the others serve as backups.

A feedthrough collar with three ports is connected to the chamber top. Two of the ports hold electrical feedthroughs for target temperature control and the beam focusing detector, and one port provides the fluid feedthrough. The electrical feedthrough for the beam focusing detector provides both low voltage to the SiPMs and signal readout for each of the three SiPMs. The fluid feedthrough contains the stainless steel tubing for hydrogen gas inlet and outlet.

\subsection{Vacuum chamber}

The vacuum chamber, shown in the left panel of Fig.~\ref{Vac_cham}, is 66\,cm high~\cite{Creare}. Due to the considerable space constraints imposed on the target system by the detectors, its horizontal dimensions are limited to a circular area with a diameter of 48\,cm. The chamber is made of stainless steel plates with a thickness of 9.5\,mm. Based on structural stress analysis using the ANSYS simulation software, this wall thickness provides at least a factor of two safety factor from the yield point. The chamber has a trapezoidal shape so that the two large side exit windows are both flat\footnote{Although it was preferred to have a vacuum chamber with a single, continuous exit window that covers the entire $\theta$ scattering angle range between $-100^\circ$ and $+100^\circ$, no solution was obtained that did not produce pleats in the vacuum window material in the middle of the acceptance. Those pleats lead to inhomogeneities which are difficult to simulate and to large multiple scattering for particles passing through them.} and parallel to the detector planes.

The beam entrance window has a $7$\,cm diameter clear aperture for the $2$\,cm FWHM beam. It is made of a $125\,\mu$m-thick aluminized Kapton$^{\text \textregistered}$ sheet, a material known for its excellent radiation resistance. The two rectangular side exit windows are each $337$\,mm wide and $356$\,mm high and cover the azimuthal angle range $\theta\,\in\,[20^{\circ}, 100^{\circ}]$ on either side of the beamline. The polar angle $\phi$ coverage of each window is $[-45^\circ, 45^{\circ}]$ from the target center at $\theta =60^{\circ}$. These unusually large vacuum windows are covered with a sailcloth fabric which is made of Mylar$^{\text \textregistered}$ laminated on aramid fabric with an areal density of $368$\,g/m$^2$~\cite{DPC785}. This material is stronger than available Kapton$^{\text \textregistered}$ foil and can withstand the pressure difference between the atmospheric pressure outside and the $10^{-4}\,$mbar pressure inside the vacuum chamber. Each window assembly consists of a frame that is glued to the aramid fabric. The Mylar$^{\text \textregistered}$ side of the sailcloth seals onto the vacuum chamber with an o-ring. GEANT4 simulations of the multiple scattering introduced by the window material have shown an angular resolution of 18.2\,mrad, just below the experimental requirement of 19\,mrad. Exposure of the window material to the 120\,GeV/c proton beam at Fermilab has demonstrated that it is radiation resistant to at least a factor of $10$  more than the total integrated particle flux expected in MUSE.\footnote{The total integrated particle flux for MUSE is expected to be 3\,MHz $\times $ 1\,year $= 10^{14}$ particles.} The beam exit window, which is $78$\,mm wide and $356$\,mm high, is much narrower than the two side exit windows and therefore suitable for $125\,\mu$m-thick aluminized Kapton$^{\text \textregistered}$ foil. Its frame was designed to strengthen the vacuum chamber.

Due to the space constraints, a single port on the bottom of the vacuum chamber acts as the pumping port and the view port for the camera system that monitors the horizontal position of the target ladder. The vacuum chamber is mounted on a three-legged stand made of 80/20 T-slot aluminum extrusions. The stand is bolted to the platform that supports the entire MUSE experiment, and has translation mechanisms to adjust the chamber position in all three directions. %The height of the stand is chosen such that the adjustment screws lie above the detector support table.

\subsection{Lifting assembly}

The target system has a lifting mechanism, shown in the right panel of Fig.~\ref{Vac_cham}, to switch between the LH$_2$ and empty target cells about twice a day without suffering the time delay of filling and emptying the cell at that frequency. It also eliminates the need to worry about changes in density due to orthohydrogen to parahydrogen conversion\cite{LH2-density}, which can take several days\cite{ortho-para-conversion}.

\begin{figure*}[ht!]
	\centering
	\includegraphics[width=0.95\textwidth]{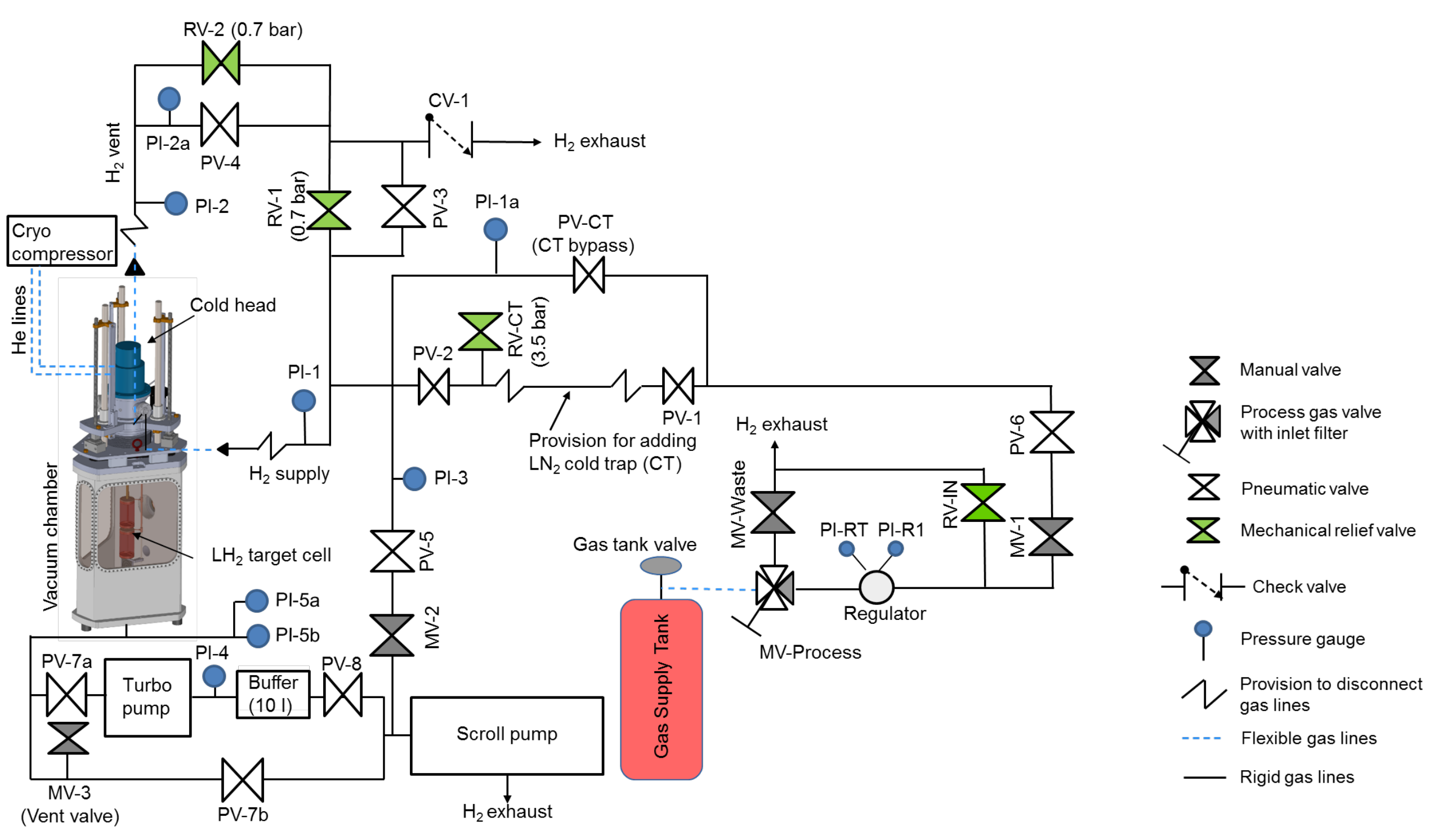}
	\caption{A schematic layout of the gas handling and vacuum systems is shown, with the legend of the symbols on the right. The flexible lines accommodate vertical movement of the ladder. The gauge pressures of the relief valves RV-1, RV-2 and RV-CT, that serve as passive safety valves for the gas system, are indicated in the diagram. Note, that RV-IN is not a primary safety valve.}
	\label{gas_flow}
\end{figure*}

The vertical movement is achieved by using three lead screws which are driven by a cogged timing belt and stepper motor system. Three guide rods keep the lifting plate centered and restrict movement to the vertical direction. In order to obtain a true and independent vertical location of the lifting plate, a magnetic position scale is attached to each guide rod at three equally spaced radial positions. A large, edge-welded bellows~\cite{bellows} with an inner diameter of 15.2\,cm accommodates the required vertical movement of 34\,cm to attain all target positions.
\subsection{Gas handling and vacuum system}

A schematic layout of the gas flow is shown in Fig.~\ref{gas_flow}. A 10\,liter gas bottle containing about 1,000\,gas liters of compressed hydrogen supplies 99.999\% pure hydrogen gas to the target. The gas bottle is placed just outside the PiM1 experimental area. The gas flow is controlled using pneumatic valves which are operated via a slow control system. Safety valves PV-3 and PV-4, which are normally-open type pneumatic valves, are set to open at $0.4$\,bar above the operating pressure of $1.1$\,bar to ensure safe venting of the target cell in case of accidental loss of compressed air or an electric power failure. Overpressure mechanical relief valves RV-1 and RV-2, which act as backup valves for the pneumatic valves, are connected in parallel to the safety valves and operate independently of the slow control system. They are set to open if the pressure in the supply or the vent lines exceeds $0.7$\,bar above the operating pressure.

The supply line and target are purged with hydrogen before starting a cooldown. The purge process involves filling the target with hydrogen gas to $1.1$\,bar (absolute) while the chamber is under vacuum, then pumping it out using a scroll pump. Once the purge process is completed, approximately 310\,liters of hydrogen gas are needed to fill the LH$_2$ target. Safe disposal of the hydrogen gas from the target cell is achieved by releasing the gas through a hydrogen vent line which sends the gas above the PiM1 area and towards the top of the experimental hall. There is no need to use a dedicated hydrogen exhaust line that directs the gas outside of the experimental hall because the approximate 0.3\,m$^3$ of hydrogen gas released into the experimental hall is much smaller than the huge volume of the experimental hall.

The vacuum system consists of scroll and turbo pumps with a pumping speed of 60\,l/s, as well as valves and a vacuum control system that is based on the Simatic HMI system from Siemens. The layout of the pumps and valves is shown in Fig.~\ref{gas_flow}. The turbo pump is backed by a scroll pump with a buffer volume to increase the lifetime of the scroll pump. This is achieved by turning off the scroll pump while the buffer volume is being filled to a pressure of 0.15\,mbar. Once that pressure is reached, the scroll pump turns back on and empties the buffer volume. This procedure reduces operation of the scroll pump by about 90\%. The same scroll pump is also used to pump out the supply line, and the target cell for purging. To avoid triggering a false fire alarm, the exhaust outlet of the scroll pump is above the target-chamber chimney which houses a hydrogen gas sensor.

\subsection{Target slow control and safety systems}
\label{sec:slow_control}

\begin{figure}[ht!]
	\centering \includegraphics[width=0.75\columnwidth]{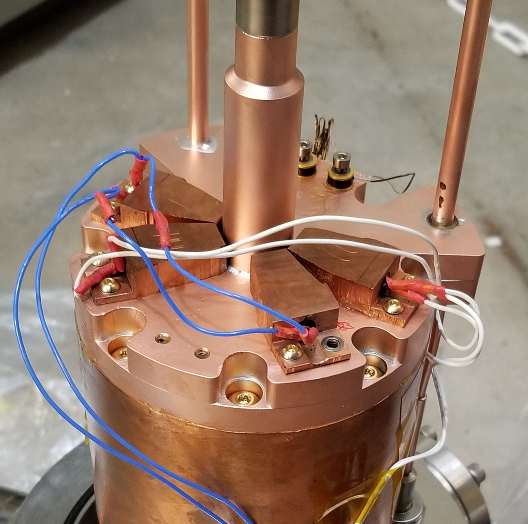}\hfill
	\caption{Photograph of two heater circuits, each consisting of two 50\,$\Omega$, 50\,W cartridge heaters connected in parallel, and two Cernox$^{\text \textregistered}$ temperature sensors attached to the bottom of the condenser, which is pictured upside down here. Each cartridge heater is glued and housed inside a copper block which is bolted to the condenser to ensure good thermal contact. }
	\label{heaters}
\end{figure}

The entire target system is monitored and controlled by a slow control system from National Instruments~\cite{slow-control} with a LabView graphical user interface. This includes the pneumatic valves of the gas system, the cryocooler compressor, the heater on the target cell, the pumping station, as well as several safety features such as interlocks and alarms. The slow control system communicates with a Lakeshore Model 336 temperature controller, which  regulates the temperature of the copper condenser, and thus of the target cell. To maintain constant temperature, the Lakeshore temperature controller reads the condenser temperature with a Cernox$^{\text \textregistered}$ temperature sensor and drives the cartridge heater circuit, shown in Fig.~\ref{heaters}, in a feedback loop. The temperature of the LH$_2$ cell is monitored by a Cernox$^{\text \textregistered}$ temperature sensor and displayed by the slow control system. The slow control system also controls the cartridge heater which is mounted on the outer face of the bottom end cap, as shown in Fig.~\ref{LH2_cell}. This heater is used to speed up the evaporation of liquid hydrogen during the target shut down process.

Hydrogen gas sensors placed in chimneys above the gas handling panel and the target chamber sound alarms if a hydrogen leak is detected. Furthermore, these sensors are interlocked to PV-6 to automatically cut-off hydrogen supply in case a leak is detected.

\section{Target Operation}
\label{sec:operation}

\begin{figure}[ht!]
	\centering \includegraphics[width=0.9\columnwidth]{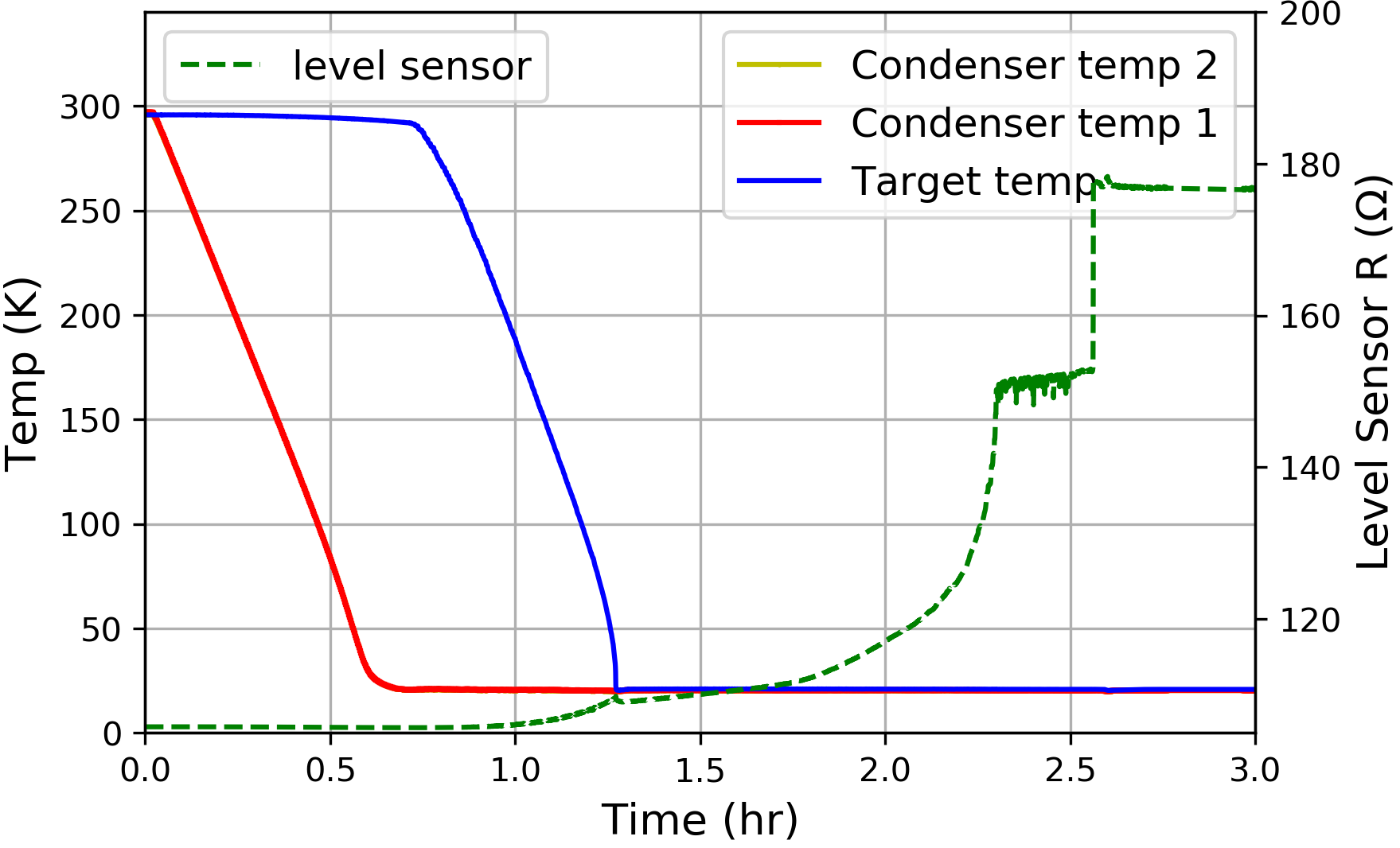}
	\caption{Results of the first hydrogen cooldown performed in the PiM1 experimental hall at PSI covering the initial 3\,hours. The red (blue)  curve shows temperature versus time for the condenser (target cell) as it is cooled down. The dashed green curve indicates the resistance of the level sensor inside the top end cap of the target cell. At about 2.3\,hours, the liquid level appears to reach the top end cap.  The target cell is full about 2.5\,hours after cooldown start.}
	\label{h2_cooldown}
\end{figure}

After passing all safety reviews, the MUSE LH$_2$ target was filled with liquid hydrogen and ran steadily for over three days. Figure~\ref{h2_cooldown} shows the first hydrogen cooldown. The red curve shows the temperature of the condenser as it is cooled down from room temperature to 21\,K in about 40 minutes. The blue curve shows the temperature of the bottom of the target cell as it is cooled down by liquid hydrogen dripping from the condenser until it reaches liquid hydrogen temperature. The dashed green curve indicates the resistance of the level sensor inside the top end cap of the target cell. The increase in the resistance is due to improved cooling of the sensor resistor as it is first in contact with and then immersed in LH$_2$. The sharp increase, followed by a constant resistance after about 2.6\,hours indicates that the target is ``full''.

\begin{figure}[ht!]
	\centering \includegraphics[width=0.9\columnwidth]{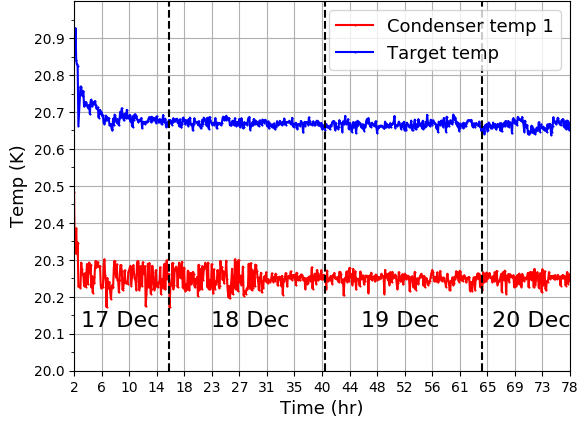}
	\caption{Results of the first hydrogen cooldown performed in the PiM1 experimental hall at PSI covering the time period from $2 - 78$\,hours. It takes about 6\,hours for the target cell temperature to stabilize. The jitter level in the condenser readout channel (red curve) was reduced significantly at hour 31 by increasing the signal readout averaging time from 1 to 5 seconds.}
	\label{h2_temp}
\end{figure}

\begin{figure}[ht!]
	\centering \includegraphics[width=0.9\columnwidth]{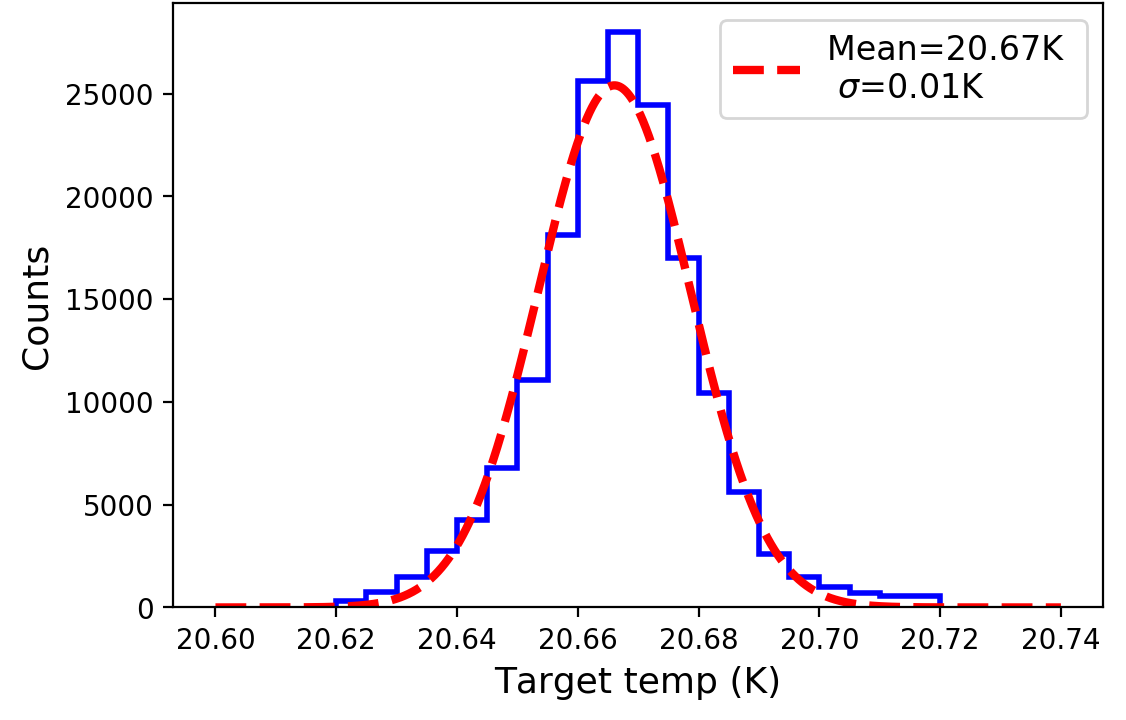}
	\caption{Histogram of target temperature values measured during hours $6-78$. The shape of the distribution is purely Gaussian with a mean of 20.67\,K and a standard deviation of 0.01\,K. }
	\label{h2_stability}
\end{figure}

Figure~\ref{h2_temp} shows data from the same cooldown as in Fig.~\ref{h2_cooldown}. It takes about 6\,hours for the target temperature to stabilize. Note that due to the low beam power and the considerable cooling power provided by the cryocooler, it is not possible to tell when beam was on or off during that time period. Figure~\ref{h2_stability} shows the variation of the target temperature at the target cell operating temperature of 20.67\,K during the time period displayed in Figure~\ref{h2_temp}. It demonstrates that the target cell temperature was constant at the $ 0.01\,$K level over the entire 72\,hour period. %Incidentally, this is equivalent to the $0.01\,$K fluctuations of the Cernox$^{\text \textregistered}$ temperature sensor~\cite{Cernox}. It is therefore possible that the real target temperature is stable even below the $0.01\,$K level.
The 20.67\,K target temperature corresponds to an operating pressure of about 1.1\,bar\footnote{The target was operated at 20.67\,K along the liquid/gas saturation line, which corresponds to a pressure of very close to 1.1\,bar. For details see https://webbook.nist.gov/chemistry/fluid/.} and a target density of 0.070\,g/cm$^3$. % that was stable to within 0.02\%.

\section{Conclusions}
\label{sec:conclusions}

A liquid hydrogen target was built for the MUSE experiment at PSI which measures the proton charge radius with high precision via both elastic muon-proton and electron-proton scattering. In order to take advantage of the unique opportunities afforded by the PiM1 beamline at PSI, and to control for systematic uncertainties at the sub-percent level, strong constraints were put on the material budget around the target and on its temperature stability. Data collected from a commissioning run at PSI demonstrated that the target was run stably at a temperature of $(20.67\pm0.01)$\,K over several days. This temperature stability suggests a liquid hydrogen density stability of 0.02\% once equilibrium concentration of parahydrogen and orthohydrogen has been reached~\cite{ortho-para-conversion}.

\section*{Acknowledgments}
We would like to thank Sheldon Stokes for leading the Creare Inc. team to help us design and build the target system, and Walter Fox (Indiana University) and Dany Horovitz (consultant to Hebrew University of Jerusalem) for their initial conceptual designs. We are grateful to Ben van den Brandt (PSI) for his many useful discussions and his specific suggestions for how to build the Kapton$^{\text \textregistered}$ target cells. We also thank Klaus Kirch, Malte Hildebrandt and Stefan Ritt (all PSI) for their their help and feedback with the target safety procedures and documentation, and the PSI vacuum group for providing the vacuum system, and especially Pascal Mayer for his design and assembly of the system. We acknowledge Manuel Schwarz and Thomas Rauber (both PSI) for their continuous help and support, and Steffen Strauch (University of South Carolina) for his help with target simulations. W.J.~Briscoe gratefully acknowledges the helpful conversations with Greg Smith (JLab), Kurt Hanson (MAXlab) and Andreas Thomas (MAMI) during the target R\&D stage. This work was supported in part by US National Science Foundation grants 1614456, 1614850, 1614938, 1649873 and 1807338. The initial conceptual design was supported by US Department of Energy grant DE-SC0012485.


\begin{thebibliography}{10}
\expandafter\ifx\csname url\endcsname\relax
  \def\url#1{\texttt{#1}}\fi
\expandafter\ifx\csname urlprefix\endcsname\relax\def\urlprefix{URL }\fi

\bibitem{CODATA2006}
P.J.~Mohr, B.N.~Taylor, and D.B.~Newell, {CODATA Recommended Values of the Fundamental Physical Constants: 2006}, Rev. Mod. Phys. \textbf{80}, (2008) 633.

\bibitem{Pohl2010}
R.~Pohl et al., {The size of the proton}, Nature \textbf{466}, (2010) 213.

\bibitem{protonpuzzle-2013}
R.~Pohl, R.~Gilman, G.A.~Miller, and K.~Pachucki, {Muonic hydrogen and the proton radius puzzle}, Annu. Rev. Nucl. Part. Sci. \textbf{63}, (2013) 175.

\bibitem{MUSE_TDR}
R.~Gilman at al., {Technical Design Report for the Paul Scherrer Institute Experiment R-12-01.1: Studying the Proton ``Radius'' Puzzle with $\mu p$ Elastic Scattering}, arXiv:1709.09753v1 [physics.ins-det].

\bibitem{Creare}
The vacuum Chamber was designed by Creare Inc. and built by Sharon Vacuum Co Inc.

\bibitem{BFD}
The scintillators are made of $2\times 2\times 2\,$mm$^3$ Saint-Gobain BC404 plastic and are connected by 3\,mm-diameter Saint-Gobain BCF-98 SC light guides to Hamamatsu S13360-3050PE silicon photomultipliers. Each SiPM is operating at 55.5\,V, consuming less than $0.2\,\mu$A; this means the total power is below $33.3\,\mu$W.

\bibitem{DPC785}
The two large vacuum windows are covered with a sailcloth fabric by Dimension-Polyant, called D-P C785,  which is made of Mylar$^{\text \textregistered}$ laminated on aramid fabric with an areal density of $368$\,g/m$^2$.

\bibitem{bellows}
The large, edge-welded bellows was manufactured by Metal-Flex$^{\text \textregistered}$ Inc. It has part number 75060-9W/FLGES with an outer diameter of 19.1\,cm and an axial stroke of 30.5\,cm.

\bibitem{slow-control}
The slow control system is based on the Crio-9035 system from National Instruments. All safety related tasks are run on a built-in FPGA. The graphical interface is based on the LabVIEW Real-Time Module from National Instruments.

\bibitem{LH2-density}
J.W.~Leachman, R.T.~Jacobsen, S.G.~Penoncello, and E.W.~Lemmon, {Fundamental Equations of State for Parahydrogen, Normal Hydrogen, and Orthohydrogen}, J. Phys. Chem. Ref. Data \textbf{38}, (2009) 721.
%https://doi.org/10.1063/1.3160306

\bibitem{ortho-para-conversion}
The saturated liquid density variation between parahydrogen and orthohydrogen at 20\,K is about 0.1\%. The equilibrium concentration of orthohydrogen and parahydrogen changes from room temperature, where it is nearly 75\% orthohydrogen and 25\% parahydrogen, to above 99\% parahydrogen and below 1\% orthohydrogen at 20\,K~\cite{LH2-density}. Note that it can take several days to achieve equilibrium.

\end{thebibliography}
\end{document}